\documentstyle[12pt,prl,aps,epsfig,citesort,psfig19,floats]{revtex}

\def\eps{{\epsilon}}

\def\begineq{\begin{equation}}
\def\endeq{\end{equation}}
\def\be{\begin{equation}}
\def\ee{\end{equation}}

\begin{document}
\bibliographystyle{prsty}

\title{
Scaling of global momentum transport  \\  
in Taylor-Couette and pipe flow
}
\author{
Bruno Eckhardt$^1$,
Siegfried Grossmann$^1$, and
Detlef Lohse$^2$
}
\address{
$^1$ Fachbereich Physik, Philipps-Universit\"at Marburg,
Renthof 6, D-35032 Marburg, Germany\\
$^2$
Department of Applied Physics  and J.\ M.\ Burgers Centre for
Fluid Dynamics, 
University of Twente, 7500 AE Enschede, 
Netherlands\\
}

\date{\today}

\maketitle
\begin{abstract}
We interpret measurements of the Reynolds number dependence
of the torque in Taylor-Couette flow by 
Lewis and Swinney [Phys. Rev. E 59, 5457 (1999)] and 
of the pressure drop in pipe flow
by Smits and Zagarola, [Phys. Fluids 10, 1045 (1998)]
within the scaling theory of Grossmann and Lohse
[J. Fluid Mech.\ 407, 27 (2000)], developed in the context of 
thermal convection. The main idea is to split the energy dissipation
into contributions from a boundary layer and the turbulent bulk. 
This ansatz can account for the observed scaling in both cases
%
if it is assumed that the internal wind velocity $U_w$ introduced
through the rotational or pressure forcing
is related to the 
the external (imposed)
velocity $U$, by $U_w/U \propto Re^\xi$ with 
$\xi = -0.051$ and 
$\xi = -0.041$ for the Taylor-Couette ($U$ inner cylinder velocity) and 
pipe flow ($U$ mean flow velocity) case, respectively.
In contrast to the Rayleigh-B\'enard case the
scaling exponents cannot (yet) be derived from the dynamical equations.
\\
\noindent Pacs: 47.27.-i 
\end{abstract}


\section{Introduction}

The relation between global flow properties and driving
forces is interesting from a fundamental point of view 
and for
upscaling from laboratory experiments to applications.
Examples are
the change in mean flow through a pipe as a function of pressure drop, 
the dependence of the heat transport 
as a function of temperature difference (Rayleigh-Benard (RB) flow), and
the increase in torque required to maintain a certain rotation speed 
in a Taylor-Couette (TC) system. 
In all three systems the
effects of the boundary layers are of prime 
importance but the way in which they are dealt with 
differs. For pipe flow the boundary
effects are usually discussed in terms of the 
Prandtl-van Karman theory \cite{ll87} which assumes 
a logarithmic law for the profile.
{\footnote{Barenblatt and coworkers however also 
succeeded to describe the data in terms of power laws  \cite{barenblatt}.}} 
For the connection between mean flow and
pressure drop it predicts an implicit {\it logarithmic} relationship,
the so called skin friction law \cite{ll87}, in reasonable agreement
with the experimental data.

On the other hand, Rayleigh-B\'enard (RB) convection
has mainly been discussed in terms of {\it algebraic} relations:
In the last decade the power law relation $Nu \sim Ra^{2/7}$ 
between the Nusselt number $Nu$ and the Rayleigh number $Ra$ was
thought to be an appropriate description of the experimental data
\cite{cas89,sig94}. Recently,
it has turned out that 
the dependences are more involved \cite{cha97,nie00,ahl00}.
On the theoretical side, 
the analysis by Grossmann and Lohse \cite{gro00}
of the different dominant dissipation mechanisms 
leads to a detailed phase diagram
for Rayleigh-B\'enard convection that is in good agreement
with the latest \cite{nie00,ahl00} and older 
experiments. In this theory the relations between $Nu$ and
$Ra$ are again {\it algebraic}.

For the Taylor-Couette system with rotating inner cylinder and resting 
outer one a description of the relation between
the dimensionless torque $G$ and the Reynolds number $Re$ both 
in terms of a skin friction law and a pure power law has been tried
\cite{lew99,lat92a}.
The dimensionless torque $G= T/\rho \nu^2 L$ 
and the Reynolds number $Re= \Omega a(b-a)/\nu$ are defined with
$T$ the torque,
$\rho$ the fluid density and $\nu$ its kinematic viscosity,
$L$ the length of the cylinders, 
$\Omega$ the angular rotation rate, and $b$ and $a$ the radii of the
outer and inner cylinder, respectively. 
Lewis and Swinney's analysis \cite{lew99} of their experimental data 
clearly shows that a pure power law
$G\sim Re^\alpha$
with $\alpha = 5/3$ as suggested in Refs.~\cite{bar84,kin84} 
does not describe the data. A description in terms of the
skin friction law is in better agreement with the data. 
However, it still
is not fully satisfactory, either, as the systematic drifts
in their Fig. 1 show. 

It is our aim here to adopt Grossmann and 
Lohse's Rayleigh-B\'enard theory to Taylor-Couette and pipe flow.
In contrast to the RB case, it is presently not possible  to 
derive the scaling exponents fully from dynamical equations. 
Instead, there will be one exponent that has to be and
can be fitted consistently to data for both systems.


We begin in section 2 with a discussion of the theory and the comparison
to Lewis and Swinney's \cite{lew99} experimental data for TC flow.
In section 3 we will adopt it to pipe flow for which precise high
$Re$ data on the pressure drop were obtained by Smits's group \cite{smi98}.
In section 4 we compare the quality of the data fit of this theory with
that of the standard skin friction law theory. 
Section 5 gives conclusions.

\begin{figure}[htb]
\setlength{\unitlength}{1.0cm}
\begin{picture}(6,8.5)
\put(-0.0,8.0){a)}
\put(8.9,8.0){b)}
\put(0.0,8.0)
{\psfig{figure=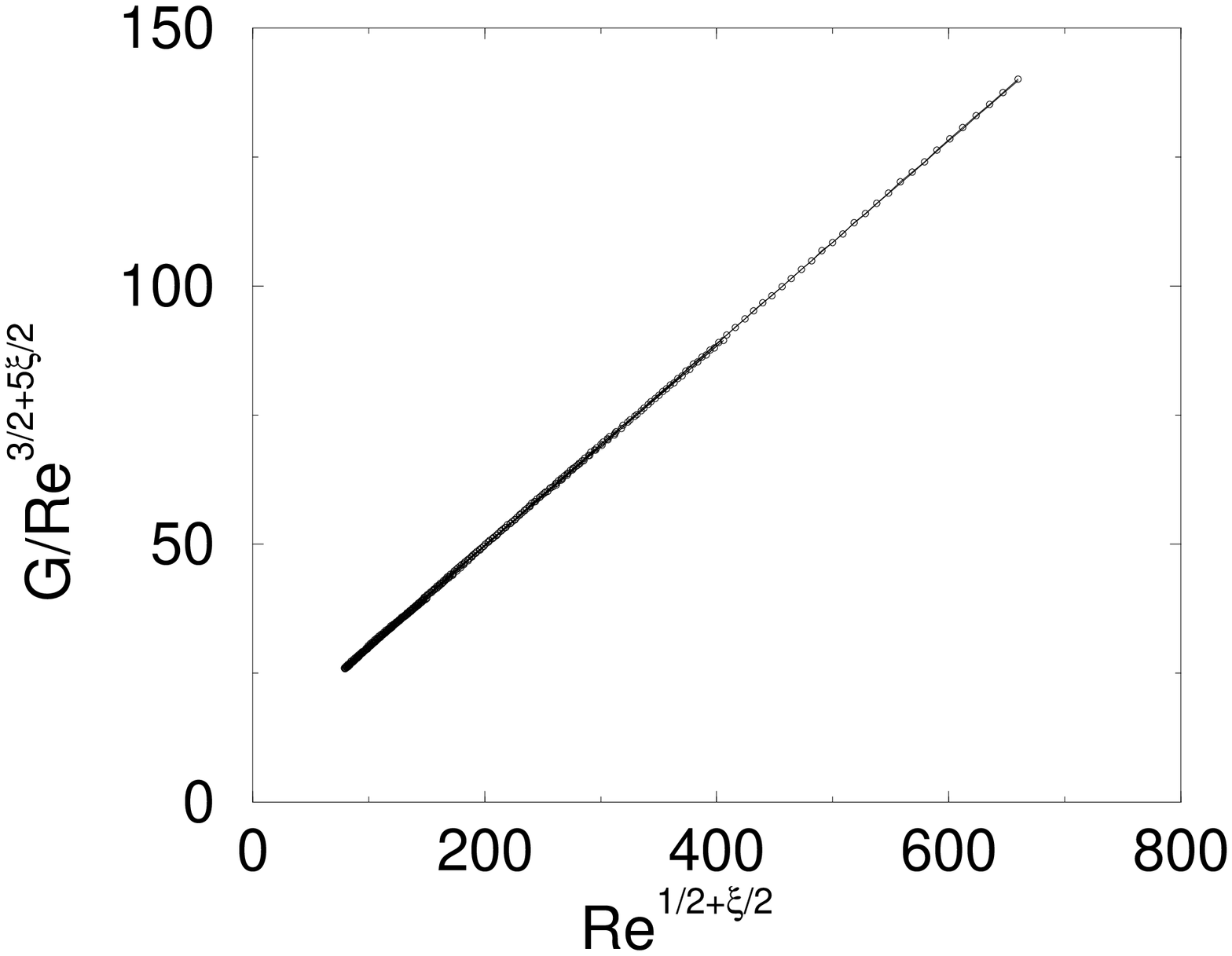,width=7cm,angle=-90}}
\put(8.9,8.0)
{\psfig{figure=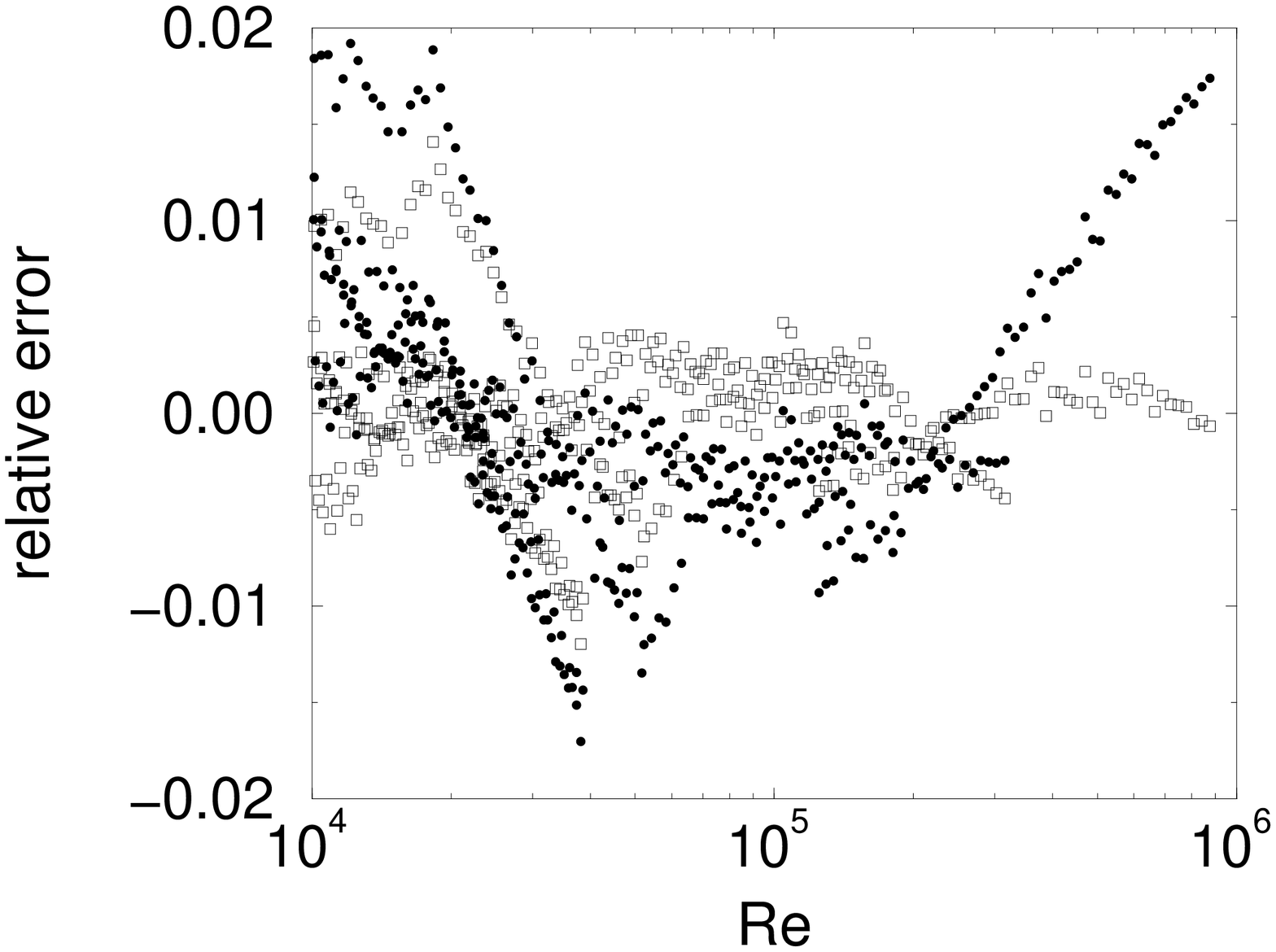,width=7cm,angle=-90}}
\end{picture}
\caption[]{
a) Compensated plot $G/Re^{3/2+5\xi/2}$ vs $Re^{1/2+\xi/2}$ with 
$\xi=-0.051$. The points are  Lewis and Swinney's data \cite{lew99}, the line
the fit  (\ref{mainprime}). 
$Re$ varies in the range $10^4$ through $9\cdot10^5$. \\
b) Relative error $(G-G_{fit})/G$ of the suggested
combination of power laws (\ref{mainprime}) with $\xi=-0.051$ (open boxes)
and relative error 
$(f-f_{fit})/f$ of the friction law fit (\ref{fric_law}) 
(filled circles).
}
\label{tc_neu}
\end{figure}

\section{Taylor-Couette flow}
The basic idea \cite{gro00}
behind the analysis of the thermal convection experiments
is the splitting of the 
 energy dissipation $\epsilon$ into contributions from the boundary layer
 (BL) and the bulk,
\be
\eps = \eps_{BL} + \eps_{bulk}. 
\label{decomp}
\ee
For the energy dissipation $\eps$ in TC flow one strictly has \cite{lew99} 
\be
\eps = {\nu^2 G\Omega \over 2\pi (b^2 - a^2)}.
\label{exact}
\ee
This can be derived as usual by considering the energy balance in a
Navier-Stokes flow.
In  analogy to ref.\ \cite{gro00} (see there for an
extensive discussion) 
$\eps_{bulk}$ is  estimated as
\be
\eps_{bulk}\sim  {U_w^3\over b-a }.
\label{epsbulk}
\ee
Here, $U_w$ is the typical velocity difference between
the turbulent and the laminar (linear) profile. It is  a measure
for the turbulent activity induced by the rotation of the inner
cylinder and it defines a Reynolds number $Re_w = U_w (b-a)/\nu$.
In the laminar case $U_w=Re_w=0$, and therefore $\eps_{bulk} = 0$.
Obviously, $U_w$ must not be confused with 
the velocity $ U = 2\pi a \Omega$ of the inner cylinder or the
corresponding Reynolds number $Re= \Omega a (b-a)/\nu$. 
The Reynolds number $Re$ (or $U$) is {\it imposed} on the flow
whereas $Re_w$ (or $U_w$) is the {\it response} of the system.
The situation can be compared with RB convection where the
Rayleigh number is imposed on the cell whereas the response of the system
is the large scale wind of turbulence, which again defines a wind Reynolds
number $Re_w$. Therefore, in analogy, also here
we call $U_w$ the wind velocity.

It is this wind velocity which leads to the formation of a boundary 
layer of thickness 
$\lambda_u$. As in  ref.\ \cite{gro00}
we assume the BL to be of Blasius 
type \cite{ll87},
\be
\lambda_u \sim (b-a) /\sqrt{Re_w}. 
\label{blasius}
\ee
For very small $Re_w$ the BL will of course not diverge
but saturate at a scale $\lambda_u \sim (b-a)$ which
introduces different scaling relations \cite{gro01}, but 
in the present work we are not interested in this very low
Reynolds number regime.

With its thickness $\lambda_u$ as 
the relevant length scale we estimate the energy dissipation 
in the BL as \cite{gro00}
\be
\eps_{BL}\sim  \nu {U_w^2 \over \lambda_u^2 } {\lambda_u \over b-a}.
\label{epsbl}
\ee

Putting eqs.\ (\ref{exact}) to (\ref{epsbl}) together one obtains
\be
G Re = c_1 Re_w^{5/2} + c_2 Re_w^3, 
\label{main}
\ee
where $c_1$ and $c_2$ are two unknown constants.
The first term is the BL contribution, the second one
the bulk contribution. 

The central question now is: How does $Re_w$ 
depend on $Re$? We do not know, but for large enough $Re$
it seems reasonable to assume a power law dependence,
\be
{Re_w \over Re } \sim 
{U_{w}\over U} \sim Re^\xi .
\label{uc}
\ee
Therefore,
\be
G= c_1 Re^{3/2+5\xi/2} + c_2 Re^{2+3\xi} . 
\label{mainprime}
\ee
We perform a nonlinear fit of  Lewis and Swinney's data
 \cite{lew99} to eq.\ (\ref{mainprime}), obtaining 
$\xi=-0.051$, $c_1=10.5$, and $c_2=0.196$.
The best way to check the quality of the 
fit  
(\ref{mainprime}) is to plot
$G/Re^{3/2+5\xi/2}$ vs $Re^{1/2+\xi/2}$ so
that according to eq.\ (\ref{mainprime}) a straight line
should result. This indeed is the case, as shown in 
Fig.~\ref{tc_neu}a. The quality of the fit is underlined
by the relative error shown in 
Fig.~\ref{tc_neu}b.

\begin{figure}[htb]
\setlength{\unitlength}{1.0cm}
\begin{picture}(6,8.5)
\put(-0.0,8.0){a)}
\put(8.9,8.0){b)}
\put(-0.0,8.0)
{\psfig{figure=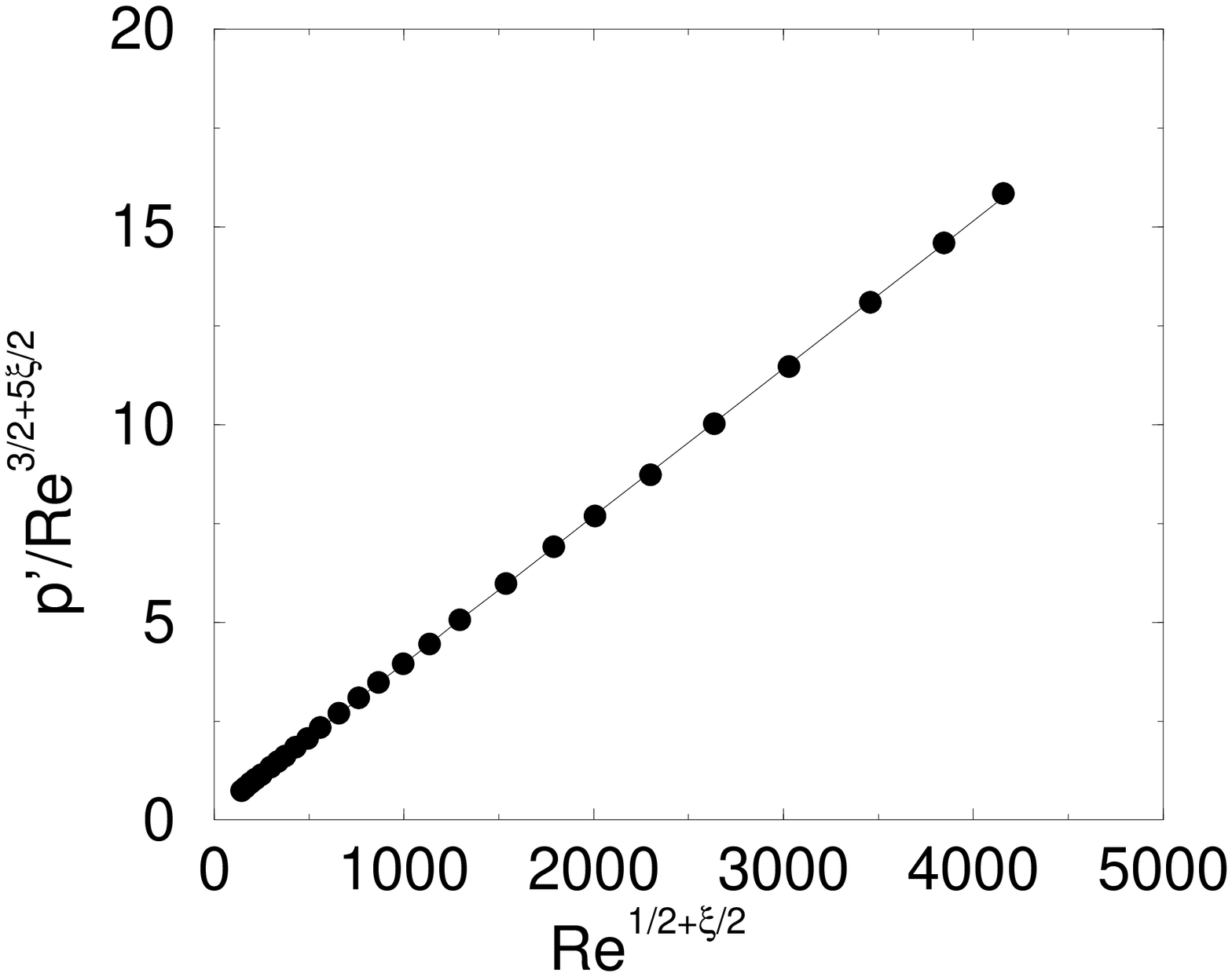,width=7cm,angle=-90}}
\put(8.9,8.0)
{\psfig{figure=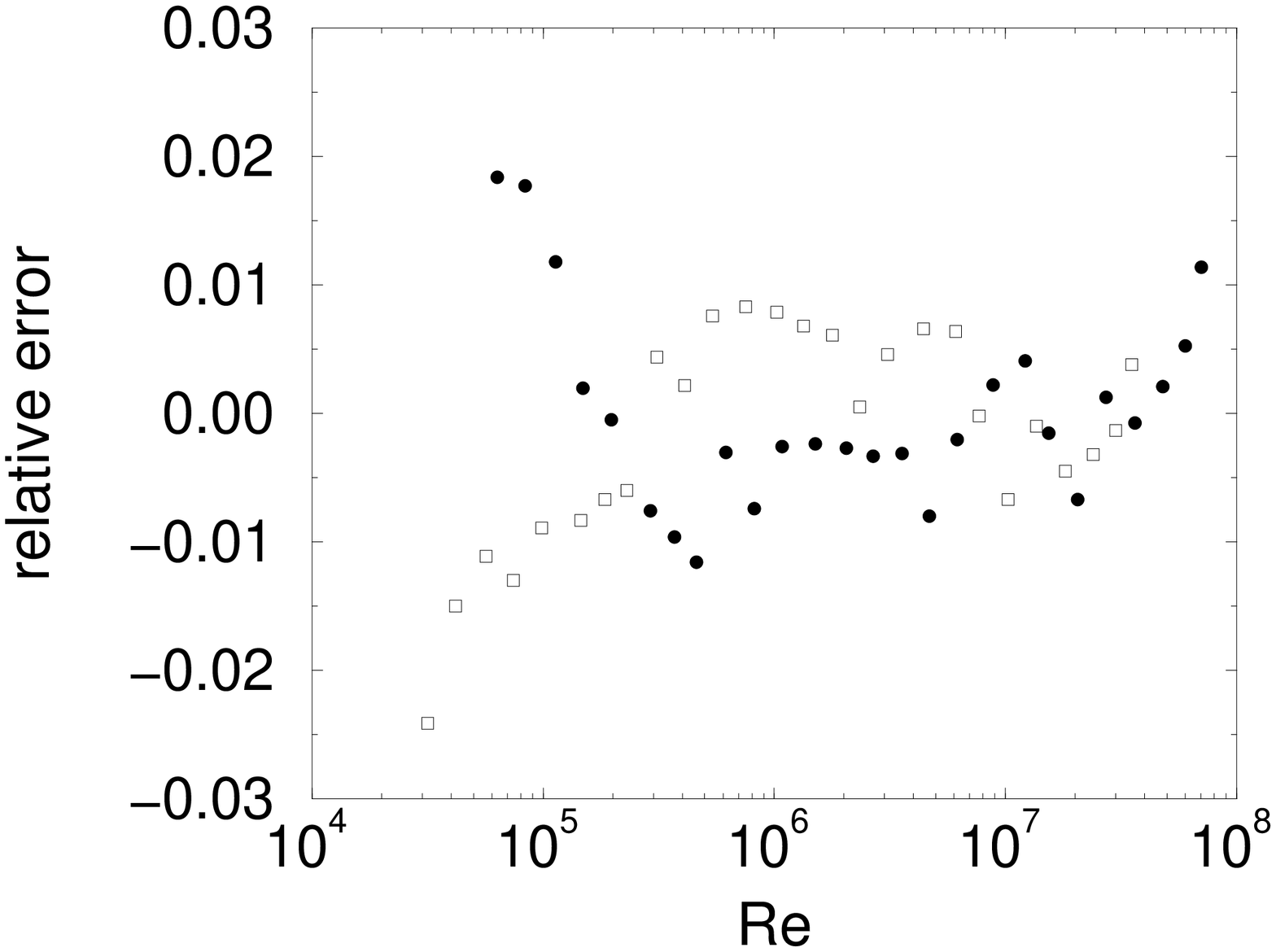,width=7cm,angle=-90}}
\end{picture}
\caption[]{
a)
Compensated plot $p^\prime/Re^{3/2+5\xi/2}$ vs $Re^{1/2+\xi/2}$ with 
$\xi=-0.041$. The points are data from ref.\ \cite{smi98}, the line
the fit  (\ref{mainprimep}). $Re$ here varies between $3.16\cdot 10^4$ and 
$3.53\cdot 10^7$. \\
b)
Relative error $(p^\prime-p^\prime_{fit})/p^\prime$ 
of the suggested
combination of power laws (\ref{mainprimep}) 
with $\xi=-0.041$ (open boxes) and 
relative error $(f-f_{fit})/f$ of the friction law fit (\ref{fric_law}) 
(filled circles).
}
\label{pipe}
\end{figure}

\section{Pipe flow}
Consider now pressure driven flow  
through a pipe of radius $R$.  
The pressure gradient and the energy dissipation
$\epsilon$ per volume are related by
\be
\eps = 
\frac{\Delta p}{\rho} \overline{u_x} 
\label{basics}
\ee
where $\overline{u_x}=:U$ is the average of the $x$-velocity 
over the cross section of the pipe, which defines the Reynolds
number $Re=2R U/\nu$.
As for the TC flow we define the wind velocity $U_w$ as the maximal difference
between the turbulent mean velocity profile and 
the laminar (parabolic) one. This maximum 
will occur close to the walls.
$U_w$ again defines a wind Reynolds number $Re_w = 2R U_w /\nu$.

We split the energy dissipation in a boundary layer
part and a bulk part as in equation (\ref{decomp}) 
and estimate both contributions as above, eqs.\ 
(\ref{epsbulk}) and (\ref{epsbl}), with $b-a$ replaced by $R$,
and the thickness of the Blasius boundary layer being
$\lambda_u \sim R/\sqrt{Re_w}$.
With this the equation for the dimensionless pressure
drop $p' = \Delta p R^3 /(\rho \nu^2 )$ becomes
\be
p'= c_1' Re^{3/2+5\xi/2} + c_2' Re^{2+3\xi},
\label{mainprimep}
\ee
where $c_1'$ and $c_2'$ are  two unknown constants and $\xi$ the
power law exponent of eq.\ (\ref{uc}). Equation (\ref{mainprimep}) 
is the analog of equation~(8) in the TC case.

In order to test   equation
(\ref{mainprimep})
we 
consider the high precision pressure drop data of 
Smits and Zagarola \cite{smi98}. A nonlinear fit then results
in $\xi=-0.041$ and $c_1'=0.226$ and $c_2'=0.00373$. 
In Fig.\ \ref{pipe}a  we show 
$p'/Re^{3/2+5\xi/2}$ vs $Re^{1/2+\xi/2}$. If eq.\ (\ref{mainprimep}) 
holds, a straight line
should result which is the case. Again the quality of the fit is
demonstrated with the relative errors in 
Fig.~\ref{pipe}b.

\section{The skin friction law}
Finally we would like to compare our description of the data 
with the  standard skin friction law \cite{ll87}. 
Define the friction coefficient 
$f=G/Re^2$ for TC flow and $f=p'/Re^2$ for pipe flow. 
One then has, 
in both cases \cite{ll87,lat92a,lew99},
\be
{1\over \sqrt{f}} = c_1^{\prime\prime} lg(Re\sqrt{f}) +c_2^{\prime\prime}
\label{fric_law}
\ee
with two flow dependent constants 
$c_1^{\prime\prime}$ and $c_2^{\prime\prime}$ which can 
be connected to the von Karman constant \cite{lew99}.
Employing equation (\ref{fric_law}) one can indeed fit both
data sets reasonably. We do not show the fits as they have already
been shown elsewhere (see figure 4a of ref.\ \cite{lew99} for the
TC case), but we present the relative error 
$(f-f_{fit})/f$ in  figures \ref{tc_neu}a and \ref{pipe}b and compare
it with the relative errors of the combined power laws eqs.\ 
(\ref{mainprime}) and (\ref{mainprimep}), respectively. 
For both TC and  pipe flow the relative error 
of the friciton law is roughly of the same
order as that of the fits (\ref{mainprime}) and (\ref{mainprimep}),
respectively, 
maybe somewhat larger.

\section{Conclusions}
The preceeding analysis shows that
the splitting of the dissipation into a bulk and a boundary layer
contribution as used in the Rayleigh-B\'enard theory 
can also be used to describe 
the Taylor-Couette flow \cite{lew99} and the pipe flow \cite{smi98} data. 
However, in contrast to the RB case, only one global balance
equation is available, the one for the energy dissipation.
Therefore, it is not possible to derive the asymptotic 
scaling exponents for both the wind Reynolds number and the
dimensionless torque (dimensionless pressure drop) in the
TC case (pipe case). Instead, one scaling exponent ($\xi$) must
be fitted to the data.
The ratio $U_{w}/U \sim Re^\xi$ scales similarly in both cases,
$\xi =-0.051$ and $\xi=-0.041$ for TC and pipe flow, respectively.
The relative error in both cases is less than one percent and
within this precession the data indicate really different exponents.
The origin of this difference is unclear.

For shear flow, the strict upper bound for energy dissipation 
is $c_\eps = \eps L/U^3 
\le 0.01087$ for $Re\to \infty$ 
\cite{bus78,nic98a}. All 
the Couette experiments clearly lie below
this bound, and even show a trend towards a scaling that is slower
than $U^3$. When assuming the Kolmogorov 
length scale as smallest length scale on which dissipation
contributes, Nicodemus et al. \cite{nic99} could numerically
show that around $Re=10^5-10^6$ one has $c_\eps \sim Re^{-0.08}$. 
Kerswell assumes the same cutoff and a certain 
background flow profile and finds   $c_\eps \sim Re^{-1/7}$
for $Re\to \infty$ \cite{ker00}. 
Interestingly enough, the exponents $3\xi = -0.12$ (TC) and $3\xi = -0.15$
(pipe) that we find from the experimental
data  are  very close to that value. However, the
scaling $U_w/U\sim Re^{\xi}$ with negative $\xi$ implies that the wind
does not increase as rapidly as the external velocity with $Re$ 
and that according to our
definition of $U_w$ the relative difference between 
the laminar
and the mean turbulent velocity profiles vanishes. 
Given the smallness of 
$\xi$ the Reynolds numbers at which this could become significant
are not experimentally accessible. But the situation remains
unsatisfactory and it clearly would be
highly desirable to calculate this exponent more rigorously from
the Navier-Stokes equations and to understand the relation to the
mean flow profile better.

\vspace{0.5cm}

\noindent
{\bf Acknowledgements:}
The authors thank Ch. Doering
and K. R. Sreenivasan   for very helpful
discussions and H. Swinney, G. Lewis, A. J. Smits, and M.~V. Zagarola
for supplying us with their experimental data. 
The work is part of the research  program of the Stichting voor 
Fundamenteel Onderzoek der Materie (FOM), which is financially supported 
by the Nederlandse  Organisatie voor Wetenschappelijk Onderzoek (NWO).
This research was also supported 
by the German-Israeli Foundation (GIF),
by the European Union (EU) under contract HPRN-CT-2000-00162,  
and in part by the National Science Foundation (NSF) 
under Grant No. PHY94-07194. We thank the members of the Institute
for Theoretical Physics in Santa Barabara for their hospitality.

\vspace{0.5cm}


\end{document}